\documentclass[10pt,pra,twocolumn,superscriptaddress,showpacs,floatfix]{revtex4}

\usepackage{graphicx}
\usepackage{amssymb}
\usepackage{times}
\usepackage{amsmath}
\usepackage{amsthm}

\begin{document}

\title{Continuous-Variable Quantum Key Distribution using Thermal States}

\author{Christian Weedbrook}\email{christian.weedbrook@gmail.com} \affiliation{Center for Quantum Information and Quantum Control, Department of Electrical and Computer Engineering and
Department of Physics, University of Toronto, Toronto, M5S 3G4, Canada}

\author{Stefano Pirandola} \affiliation{Department of Computer Science, University of York, York YO10 5DD, United Kingdom}

\author{Timothy C. Ralph} \affiliation{Centre for Quantum Computation and Communication, School of Mathematics and Physics, University
of Queensland, St Lucia, Queensland 4072, Australia}

\date{\today}

\begin{abstract}

We consider the security of continuous-variable quantum key distribution using thermal (or noisy) Gaussian resource states. Specifically, we analyze this against collective Gaussian attacks using direct and reverse reconciliation where both protocols use either homodyne or heterodyne detection. We show that in the case of direct reconciliation with heterodyne detection, an improved robustness to channel noise is achieved when large amounts of preparation noise is added, as compared to the case when no preparation noise is added. We also consider the theoretical limit of infinite preparation noise and show a secure key can still be achieved in this limit provided the channel noise is less than the preparation noise. Finally, we consider the security of quantum key distribution at various electromagnetic wavelengths and derive an upper bound related to an entanglement-breaking eavesdropping attack and discuss the feasibility of microwave quantum key distribution.

\end{abstract}


\maketitle

\section{Introduction}

Continuous-variable quantum key distribution (QKD)~\cite{Sca09,Weedbrook2011} is the ability to generate a secret key between two distant parties, Alice and Bob, which can be used to encrypt messages for secure communication. This is achieved by using Gaussian quantum resource states~\cite{Weedbrook2011} where its theoretical security is guaranteed by the no-cloning theorem. A typical Gaussian modulated protocol~\cite{Grosshans2002,Grosshans2003,Sil02,C.Weedbrook2004,Lan05,S.Pirandola2008} involves Alice randomly displacing a number of pure vacuum modes and sending them over an insecure quantum channel to Bob. These modes are then measured by Bob using either homodyne~\cite{Grosshans2002} or heterodyne detection~\cite{C.Weedbrook2004}. The various stages of classical communication~\cite{Sca09} follow next, including error correction, where either direct~\cite{Grosshans2002} or reverse reconciliation~\cite{Grosshans2003} can be used.

Generally, in Gaussian QKD protocols, it is assumed that Alice starts off with a large number of \textit{pure} vacuum states. However, this is an idealization and is never quite true in practice with small amounts of unknown Gaussian preparation noise often being present. The idea of analyzing the security of continuous-variable QKD using such noisy or thermal states was first considered in~\cite{Fil08,Usenko2010}. Here they showed using reverse reconciliation that the distance over which continuous-variable QKD was secure declined rapidly as the resource states became noisier, ultimately resulting in the inability to generate a secure key. However, in a subsequent work~\cite{Weedbrook2010}, it was shown using direct reconciliation that the distance with which the protocol is secure does not decline to zero as the states become noisier. In fact, even though the rate of generation of the secret key decreases for increasing noise, it remains bigger than zero for values of transmission $T > 0.5$. This means that the security threshold of the protocol remains at $T=0.5$ for extremely high values of preparation noise. Thus, up to a requirement of a strong modulation of the input, thermal state QKD is able to reach distances comparable to standard QKD. Furthermore, an application of the analysis of noisy coherent states was found by considering the security of QKD at various wavelengths of the electromagnetic spectrum, revealing regions of security from the optical all the way down into the microwave region~\cite{Weedbrook2010}.

In this paper, we build upon the work presented in~\cite{Weedbrook2010} and outline our results here. We begin by using the previous analysis of reverse~\cite{Usenko2010,Weedbrook2010} and direct reconciliation~\cite{Weedbrook2010} using homodyne detection and extending them both to study the case of heterodyne detection. For the case of direct reconciliation and heterodyne detection we show an improved robustness to channel noise when large amounts of preparation noise is added, as compared to the case with zero preparation noise. This effect of noise improving the performance
of QKD using direct reconciliation, was previously seen in the context of homodyne detection where pure coherent states were more robust than pure squeezed states~\cite{Navascues2005}. In~\cite{Weedbrook2010} it was shown that direct reconciliation could tolerate a thermal variance of $10^4$ times that of the pure vacuum mode and still show no deterioration in the security threshold of the protocol (albeit with a reduced key rate). Here we extend this result and show that, provided the channel noise is less than the preparation noise, the same protocol can, in principle, tolerate any amount of preparation noise, again at a cost of decreasing key rate. Finally, we consider the security of QKD at various electromagnetic wavelengths and develop an improved security bound along with an upper bound related to an entanglement-breaking eavesdropping attack.

This paper is organized as follows. Section II introduces the main concepts of thermal state QKD. In Secs.~III and IV the secret key rates for direct and reverse reconciliation using both homodyne and heterodyne detection are given. Section V considers QKD in the so-called classical limit where an infinite amount of preparation noise is added for direct reconciliation using homodyne detection. Finally, before concluding in Sec.~VIII, we look at the security of QKD at various wavelengths along with the feasibility of microwave QKD in Sec.~VII.

\section{Thermal State Quantum Key Distribution}

\begin{figure}[!ht]
\begin{center}
\includegraphics[width=8cm]{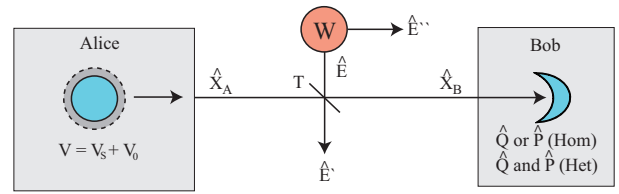}
\caption{Schematic of a continuous-variable QKD protocol using thermal states. The loss in the quantum channel is modeled by a beam splitter with channel transmission $T$. The eavesdropping attack is a Gaussian collective attack in the form of an entangling cloner attack where the variance of the EPR state is $W$ with the modes of the EPR beam described by the operators $\hat{E}''$ and $\hat{E}'$. The initial mode sent by Alice $\hat{X}_A$ is a thermal state and once Bob receives the mode $\hat{X}_B$ he will perform either a homodyne (Hom) or heterodyne (Het) measurement on it.}\label{fig_1_way}
\end{center}
\end{figure}

The initial stages of a thermal state QKD protocol consists in Alice preparing a number of randomly displaced thermal states and then sending them to Bob over an insecure quantum channel monitored by Eve (cf., Fig.~\ref{fig_1_way}). This initial mode prepared by Alice can be described in the Heisenberg picture as
\begin{align}
\hat{X}_A = X_S + \hat{X}_0,
\end{align}
where $X_S$ describes the classical signal encoding and $\hat{X}_0$ describes the quantum noise of the thermal mode. Here the quadrature variables are given by: $\hat{X}_A \in \{\hat{Q}_A, \hat{P}_A\}$, $X_S \in \{Q_A, P_A\}$ and $\hat{X}_0 \in \{\hat{Q}_0, \hat{P}_0\}$. The overall variance $V := V(\hat{X}_A)$ of Alice's initially prepared modes is given by
\begin{align}
V = V_S + V_0,
\end{align}
where $V_S$ is a Gaussian distribution with zero mean. Here $V_0$ is the shot-noise which can be defined in terms of the conditional variance as
\begin{align}
V(\hat{Q}_A|Q_A) = V(\hat{P}_A|P_A) = V_0 \geq 1,
\end{align}
where the conditional variance is defined as~\cite{Poizat1994,Grangier1998}
\begin{align}\label{eq: conditional variance def}
V(\hat{X}|Y) = V(\hat{X}) - \frac{|\langle \hat{X}Y \rangle|^2}{V(Y)}.
\end{align}
We can decompose the shot-noise variance as $V_0 = 1 + \beta$ where $\beta$ is the variance of the preparation noise at Alice's station and $1$ denotes the variance of the pure vacuum mode. It is common in most continuous-variable QKD protocols to theoretically let $V = V_S + 1$, i.e., zero preparation noise ($\beta=0$) in Alice's mode preparation. However in our analysis we consider the general case where $\beta$ is different from zero. This means that the shot-noise $V_0$ (that we also call the ``purity") can be greater than $1$. Then we make the valid assumption that this preparation noise is restricted to Alice's station and not accessible, or known, to Eve (or even to Alice for that matter).

The most important type of eavesdropping attack is the collective Gaussian
attack~\cite{Nav06,Gar06,Pir08}. It was shown that such an attack is the most powerful attack allowed by quantum physics, up to a suitable symmetrization of the protocols~\cite{Ren09}. It consists in Eve interacting her independent
ancilla modes with Alice's mode for each run of the protocol
in such a way to generate a memoryless (or one-mode) Gaussian channel.
The entangling cloner~\cite{Gross03} is the most important and practical example of a collective Gaussian attack and is used in our analysis. This consists in Eve perfectly replacing the quantum channel between Alice and Bob with her own quantum channel where the loss is simulated by a beam splitter with transmission $T \in [0,1]$.

She then prepares ancilla modes $\hat{E}$ and $\hat{E}''$ in an Einstein-Podolsky-Rosen (EPR) entangled Gaussian state~\cite{Kok2010} with variance $W$ (see Fig.~\ref{fig_1_way}). Eve keeps one mode $\hat{E}''$ and injects the other mode $\hat{E}$ into the unused port of the beam splitter, leading to the output mode $\hat{E}'$. This operation is repeated identically and independently for each signal mode sent by Alice. Eve's output modes are then stored in a quantum computer and detected collectively at the end of the protocol. Eve's final measurement is optimized based on Alice and Bob's classical communication. Note that this attack can be simply described by two parameters: the channel transmission $T$ and the channel noise $W$. The latter parameter can be replaced by the equivalent noise of the channel
\begin{align}
\chi = \frac{(1-T)}{T} + \epsilon,
\end{align}
where the first term $(1-T)/T$ corresponds to the noise induced by the loss and $\epsilon$ is the excess channel noise which can be written as $\epsilon = (W-1)(1-T)/T$. In the particular case where $W=1$, or equivalently $\epsilon=0$ (no excess noise), the attack corresponds to a pure loss channel.

\section{Reverse Reconciliation}

We begin the analysis by considering reverse reconciliation~\cite{Grosshans2003}~(denoted by the symbol $\blacktriangleleft$) using homodyne and heterodyne detection. Note that reverse reconciliation using homodyne detection has been analyzed before~\cite{Fil08,Usenko2010}. For completeness, we give the derivation for reverse reconciliation for homodyne detection so as to help in the derivation for heterodyne detection. Before commencing we make a brief comment about notation. When we consider homodyne detection the relevant variable $X_B$ is $Q_B \in \mathbb{R}$ (or $P_B \in \mathbb{R}$, equivalently). On the other hand, when considering heterodyne detection, the relevant variable $X_B$ is the pair $\{Q_B,P_B\} \in \mathbb{R}^2$. Also a variable with hat is an operator
while the same variable without a hat is the corresponding
classical variable after measurement.

\subsection{Homodyne Detection}

The secret key rate $R^{\blacktriangleleft [Hom]}$ for reverse reconciliation where Bob uses homodyne detection is given by
\begin{align}
R^{\blacktriangleleft [Hom]} := I(X_A:X_B) - I(X_B:E),
\end{align}
where the mutual information between Alice and Bob is given by
\begin{align}\label{eq: alice and bob mutual Hom RR}
I(X_A:X_B):= H(X_B) - H(X_B|X_A),
\end{align}
where
\begin{align}\label{eq: shannon entropy hom}
H(X_B) = \frac{1}{2} \log_2 V(\hat{X}_B),
\end{align}
is the Shannon (or classical) entropy and
\begin{align}\label{eq: shannon conditional_entropy hom}
H(X_B|X_A) = \frac{1}{2} \log_2 V(\hat{X}_B|X_A),
\end{align}
is the conditional Shannon entropy \cite{Sha48}. The mutual information between Eve and Bob is given by the Holevo bound \cite{Hol73} defined as
\begin{align}\label{eq: eve and bob mutual info}
I(X_B:E) := S(E) - S(E|X_B),
\end{align}
where $S(\cdot)$ is the von Neumann (or quantum) entropy. The von Neumann entropy of a Gaussian state $\rho$ containing $n$ modes can be written in terms of its symplectic eigenvalues~\cite{Hol99}
\begin{align}\label{eq: von neuman entropy}
S(\rho) = \sum_{k=1}^{n} g (\nu_k),
\end{align}
where
\begin{align}\label{eq: g of von Neumann}
g(\nu) = \Big(\frac{\nu+1}{2}\Big) {\rm log}_2 \Big(\frac{\nu+1}{2}\Big) - \Big(\frac{\nu-1}{2}\Big) {\rm log}_2 \Big(\frac{\nu-1}{2}\Big).
\end{align}
We will show how to explicitly calculate the symplectic spectrum $\boldsymbol{\nu} = \{\nu_1, ..., \nu_n\}$ soon.

To begin with though, let's calculate Alice and Bob's mutual information. To achieve this the first step is to consider the output modes at Bob's (and Eve's) station. These are given respectively by
\begin{align}\label{eq: Bob variance}
V(\hat{Q}_B) &= V(\hat{P}_B) = (1-T) W + T V := b_V,\\
V(\hat{Q}_{E'}) &= V(\hat{P}_{E'}) = (1-T) V + T W := e_V,
\end{align}
with the following conditional variances
\begin{align}\label{eq: Bob cond var}
V(\hat{Q}_B|Q_A) = V(\hat{P}_B|P_A) = (1-T) W + T V_0 := b_1,\\
V(\hat{Q}_{E'}|Q_A) = V(\hat{P}_{E'}|P_A) = (1-T) V_0 + T W := e_1,
\end{align}
derived using the definition given in Eq.~(\ref{eq: conditional variance def}). Using Eq.~(\ref{eq: alice and bob mutual Hom RR}) with Eqs.~(\ref{eq: Bob variance}) and (\ref{eq: Bob cond var}) we can calculate Alice and Bob's mutual information to be
\begin{align}\label{eq: alice and bob mutual hom RR formula}
I(X_A:X_B) = \frac{1}{2} \log_2 \Big[\frac{(1-T) W + T V_S + T V_0}{(1-T) W + T V_0}\Big].
\end{align}
Note that, ultimately in the above equation it is $V_0$ that will be varied in our calculations. Next up is the calculation of Eve and Bob's mutual information, i.e., Eq.~(\ref{eq: eve and bob mutual info}). First though we need to introduce the covariance matrix. The covariance matrix $\textbf{V}$ can be constructed using the following definitions of its matrix elements
\begin{align}\label{eq: covariance matrix def}
V_{lm}&:= \frac{1}{2} \langle \hat{Y}_l \hat{Y}_m + \hat{Y}_m \hat{Y}_l \rangle - \langle \hat{Y}_l \rangle  \langle \hat{Y}_m \rangle, \\\label{eq: CM1}
V_{ll}&= \langle \hat{Y}_l^2 \rangle - \langle \hat{Y}_l \rangle^2 := V(\hat{Y}_l),
\end{align}
where $\hat{Y}_l$ is the $l^{\rm th}$ element of the quadrature row vector $\hat{\textbf{Y}} = (\hat{Q}_1, \hat{P}_1, ..., \hat{Q}_n, \hat{P}_n)$ which describes the bosonic system of $n$ modes. As mentioned previously, to calculate Eq.~(\ref{eq: eve and bob mutual info}), we need to calculate the symplectic spectrum of the appropriate covariance matrices. The symplectic spectrum $\boldsymbol{\nu}= \{\nu_1,...,\nu_n \}$ of an arbitrary covariance matrix $\textbf{V}$ can be calculated by finding the (standard) eigenvalues of the matrix $|i {\bf \Omega} \textbf{V}|$, where ${\bf \Omega}$ defines the symplectic form and is given by
\begin{align}
{\bf \Omega}:= \bigoplus_{k=1}^{n} \left(
                                     \begin{array}{cc}
                                       0 & 1 \\
                                       -1 & 0 \\
                                     \end{array}
                                   \right).
\end{align}
Here $\bigoplus$ is the direct sum indicating adding matrices on the block diagonal.

\begin{figure}[!ht]
\begin{center}
\includegraphics[width=8cm]{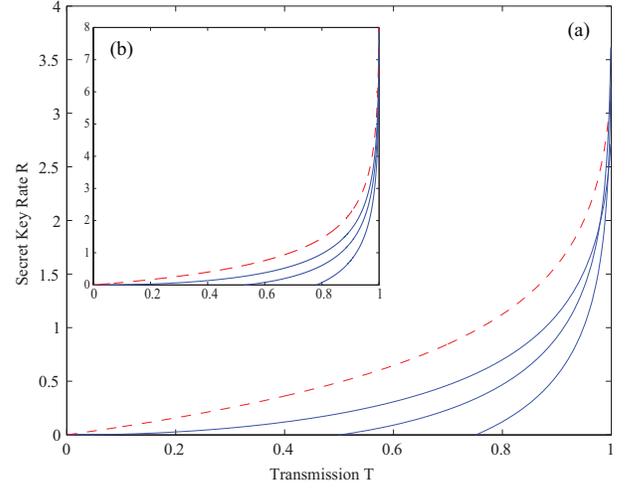}
\caption{Secret key rates for reverse reconciliation using (a) homodyne detection and (b) heterodyne detection for various values of $V_0$. Top dashed (red) line is a pure encoded state sent by Alice with the solid (blue) lines giving different values of impurity, i.e., $V_0 = 2,3,5$ from top to bottom. Here $V_S =10^3$ and $W=1$ (i.e., only loss on the quantum channel).}\label{RR_hom_all}
\end{center}
\end{figure}

Eve's covariance matrix is made up from the two modes $\hat{E}'$ and $\hat{E}''$
\begin{align}\label{eq: eves CM}
\textbf{V}_{E} (V,V) = \left(
              \begin{array}{cc}
                {\rm diag}[e_{V},e_V] & \varphi \textbf{Z} \\
                \varphi \textbf{Z} & W \textbf{I} \\
              \end{array}
            \right),
\end{align}
where $\varphi = [T(W^2 -1)]^{1/2}$ and the notation ``${\rm diag}$" simply means a matrix with the arguments on the diagonal elements of a matrix and zeros everywhere else. Here $\textbf{Z}$ and $\textbf{I}$ are the Pauli matrices
\begin{align}
\textbf{Z} = \left(
               \begin{array}{cc}
                 1 & 0 \\
                 0 & -1 \\
               \end{array}
             \right), \hspace{2mm}
\textbf{I} = \left(
               \begin{array}{cc}
                 1 & 0 \\
                 0 & 1 \\
               \end{array}
             \right).
\end{align}
To calculate Eve's symplectic spectrum we note that a  particular covariance matrix of the form
\begin{equation}
\mathbf{V}=\left(
\begin{array}{cc}
a\mathbf{I} & \sqrt{T}c\mathbf{Z} \\
\sqrt{T}c\mathbf{Z} & b\mathbf{I}%
\end{array}%
\right) :=\mathbf{V}(a,b,c,T)~,  \label{V_simple}
\end{equation}
where $c\geq 0$ and $T\in \lbrack 0,1]$, has a symplectic spectrum with a simple expression given by
\begin{equation}\label{Spectrum_simple1}
\nu _{\pm}:=\frac{1}{2}\left[ \sqrt{y} \pm (a-b) \right] ~,
\end{equation}
where $y=(a+b)^{2}-4c^{2}T\geq 4$~\cite{Weedbrook2011}. Therefore, using the above, Eve's symplectic spectrum can be expressed in a more compact form as
\begin{align}\label{eq: symplectic eigenvalues of Eve}
\nu_{E}^{\pm} = \frac{1}{2} \Big[\sqrt{(e_V+W)^2 - 4 T (W^2 -1)} \pm (e_V - W)\Big].
\end{align}
Next we need to calculate the symplectic spectrum of the covariance matrix $\textbf{V}_{E|X_B}$. This represents the covariance matrix of a system where one of the modes has been measured using homodyne detection (in this case Bob) and is given by~\cite{Eisert2002,Fiurasek2002,Weedbrook2011}
\begin{align}\label{eq: CM under homo2}
\textbf{V}_{E|X_B} = \textbf{V}_E - (b_V)^{-1} \textbf{D} {\bf \Pi}  \textbf{D}^T,
\end{align}
where
\begin{align}
{\bf \Pi}:=
\left(
  \begin{array}{cc}
    1 & 0 \\
    0 & 0 \\
  \end{array}
\right).
\end{align}
Here $\textbf{D}$ is a $4 \times 2$ matrix describing the (quantum) correlations between Eve's modes $\{\hat{E}',\hat{E}''\}$ and Bob's output mode $\hat{X}_B$. It is given by
\begin{align}\label{eq: definition of C}
\textbf{D} := \left(
               \begin{array}{c}
                 \langle \hat{E}' \hat{X}_B \rangle \textbf{I} \\
                 \langle \hat{E}'' \hat{X}_B \rangle \textbf{Z} \\
               \end{array}
             \right)
             =
             \left(
               \begin{array}{c}
                 \xi \textbf{I} \\
                 \phi \textbf{Z} \\
               \end{array}
             \right),
\end{align}
where
\begin{align}
\xi &= -\sqrt{T (1-T)} (V_S + V_0 - W),\\
\phi &= \sqrt{1-T} \sqrt{W^2-1},
\end{align}
and we have used $\hat{X}_B = \sqrt{T} \hat{X}_A + \sqrt{1-T} \hat{E}$ and $\hat{E}' = -\sqrt{1-T} \hat{X}_A +\sqrt{T} \hat{E}$. Using Eq.~(\ref{eq: CM under homo2}) we find that Eve's conditional covariance matrix is given by
\begin{align}\label{eq: block form Eve cond}
\textbf{V}_{E|X_B} =
\left(
\begin{array}{cc}
\mathbf{A} & \mathbf{C} \\
\mathbf{C}^{T} & \mathbf{B}%
\end{array}%
\right),
\end{align}
where
\begin{align}\nonumber
\mathbf{A} &=
\left(
  \begin{array}{cc}
    \frac{VW}{T (V-W) + W} & 0 \\
    0 & (1-T) V + TW \\
  \end{array}
\right),\\\nonumber
\mathbf{B} &=
\left(
  \begin{array}{cc}
    \frac{1-T + TWV}{TV + W - TW} & 0 \\
    0 & W \\
  \end{array}
\right),\\\nonumber
\mathbf{C} &=
\left(
  \begin{array}{cc}
    \sqrt{T (W^2-1)} \Big(\frac{V}{TV + W - T W}\Big) & 0 \\
    0 & -\sqrt{T (W^2-1)} \\
  \end{array}
\right).
\end{align}
%
The symplectic spectrum of the above conditional covariance matrix is composed by the two eigenvalues~\cite{Alex}
\begin{equation}\label{eq: symplectic eigenvalues def}
\nu _{\pm}=\sqrt{\frac{\Delta \pm \sqrt{\Delta ^{2}-4\det \mathbf{V}}}{2}}~,
\end{equation}
where $\det \mathbf{V}$ (the determinant of the covariance matrix) and $\Delta :=\det \mathbf{A}+\det \mathbf{B}+2\det
\mathbf{C}$ are global symplectic invariants. Note that these quantities can also be simply expressed in terms of the symplectic spectrum as
\begin{align}\label{eq: variables in terms of SS}
{\rm det} \textbf{V} = \nu^2_{+}  \nu^2_-, \hspace{2mm} \Delta = \nu^2_{+} + \nu^2_-
\end{align}
Using Eq.~(\ref{eq: symplectic eigenvalues def}) the corresponding symplectic spectrum $\boldsymbol{\nu}_{E|X_B}$ of $\textbf{V}_{E|X_B}$ can be calculated but is not written down explicitly here due to its length. Finally using Eq.~(\ref{eq: von neuman entropy}) and Eq.~(\ref{eq: g of von Neumann}) with the just computed symplectic spectra, we can determine Bob and Eve's mutual information. The final secret key rate $R^{\blacktriangleleft [Hom]}$ is calculated and plotted in Fig.~\ref{RR_hom_all}~(a) using various values of $V_0$ for a lossy channel (i.e., a quantum channel with only loss and no added noise, i.e., $W=1$). We find that, for only moderate values of $V_0$, the security of the protocol reduces rapidly.

\subsection{Heterodyne Detection}

The secret key rate for reverse reconciliation where Bob now employs heterodyne detection is given by
\begin{align}
R^{\blacktriangleleft [Het]} := I(X_A:X_B) - I(X_B:E),
\end{align}
where, as mentioned previously, $X_B$ is $\{Q_B,P_B\} \in \mathbb{R}^2$ for heterodyne detection and not $Q_B \in \mathbb{R}$ (or equivalently $P_B \in \mathbb{R}$) as it was previously for homodyne detection. The mutual information between Alice and Bob is again defined as
\begin{align}\label{eq: alice and bob mutual Het RR_def}
I(X_A:X_B):= H(X_B) - H(X_B|X_A),
\end{align}
except now the Shannon entropies are
\begin{align}\label{eq: shannon entropy het}
H(X_B) = \log_2 V(\hat{X}_B),
\end{align}
and
\begin{align}\label{eq: shannon conditional_entropy het}
H(X_B|X_A) = \log_2 V(\hat{X}_B|X_A).
\end{align}
Note that the above two formulas do not have the usual factor of $1/2$ out the front. This indicates that twice the amount of information is obtained using heterodyne detection, but at a cost of the extra unit of vacuum noise introduced at the beam splitter. The mutual information between Eve and Bob is again given by the Holevo information
\begin{align}
I(X_B:E) := S(E) - S(E|Q_B,P_B),
\end{align}
but now $S(E|Q_B,P_B)$ is calculated from the symplectic spectrum $\boldsymbol{\nu}_{E|Q_B,P_B}$ of the conditional covariance matrix $\textbf{V}_{E|Q_B,P_B}$. The variances of the quadratures of the output modes after Bob's heterodyne measurement are given by
\begin{align}
V(\hat{Q}_B) &= V(\hat{P}_B) = \frac{1}{2} (b_V + 1) := b_V',
\end{align}
where $b_V$ is defined in Eq.~(\ref{eq: Bob variance}). The following conditional variances now apply
\begin{align}
V(\hat{Q}_B|Q_A) &= V(\hat{P}_B|P_A) = \frac{1}{2} (b_1 + 1).
\end{align}
Using Eq.~(\ref{eq: alice and bob mutual Het RR_def}) we calculate Alice and Bob's mutual information to be
\begin{align}\label{eq: alice and bob mutual Het RR}
I(X_A:X_B) = \log_2 \Big[\frac{(1-T) W + T V_S + T V_0 + 1}{(1-T) W + T V_0 + 1}\Big].
\end{align}
The covariance matrix of Eve conditioned on Bob's heterodyne measurement results $\{Q_B,P_B\}$ is given by~\cite{Weedbrook2011}
\begin{align}
\textbf{V}_{E|Q_B,P_B} = \textbf{V}_E - \theta^{-1} \textbf{D} ({\bf \Omega} \textbf{V}_B {\bf \Omega^T} + \textbf{I}) \textbf{D}^T,
\end{align}
where $\textbf{V}_E$ is given by Eq.~(\ref{eq: eves CM}) and $\textbf{V}_B= b_V \textbf{I}$. Here $\theta:= {\rm det} \textbf{V}_B + {\rm Tr} \textbf{V}_B + 1$ and $\textbf{D}$ is defined previously in Eq.~(\ref{eq: definition of C}). We find that $\theta = b_V^2 + 2 b_V + 1$ and ${\bf \Omega} \textbf{V}_B {\bf \Omega}^T + \textbf{I} = \textbf{V}_B + \textbf{I}$. We find that
\begin{align}
\textbf{V}_{E|Q_B,P_B} =
\left(
\begin{array}{cc}
a\mathbf{I} & \sqrt{T} c \mathbf{Z} \\
\sqrt{T} c \mathbf{Z} & b\mathbf{I}%
\end{array}%
\right),
\end{align}
where
\begin{align}\nonumber
a &=
\frac{(1-T)V + (T+V)W}{1+TV+(1-T)W},\\\nonumber
%
%
b &=
\frac{1-T+(1+TV)W}{1+TV+(1-T)W},\\\nonumber
%
%
%
c &=
\sqrt{W^2-1} \Big[\frac{1+V}{1+TV +(1-T)W}\Big].
\end{align}
The above covariance matrix has the corresponding symplectic spectrum
\begin{align}
\nu_{E|Q_B,P_B}^{\pm} = \frac{1}{2} [\sqrt{y} \pm (a-b)]
\end{align}
where $y = (a+b)^2 -4c^2T$ as given by Eq.~(\ref{Spectrum_simple1}). Using this, the final secret key rate $R^{\blacktriangleleft [Het]}$ can be calculated and is plotted in Fig.~\ref{RR_hom_all}~(b) for different values of $V_0$.

\begin{figure}[!ht]
\begin{center}
\includegraphics[width=8cm]{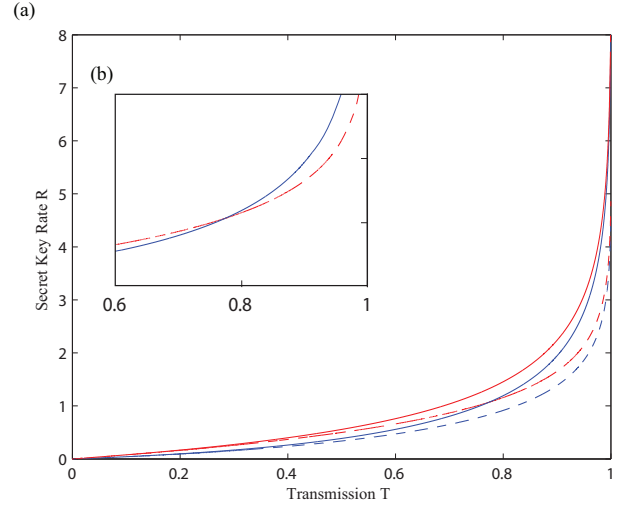}
\caption{(a) Comparison of reverse reconciliation using homodyne detection (dashed lines) against heterodyne detection (solid lines) for $V_0 = 1$ (red lines) and $V_0 = 1.5$ (blue lines) and with only loss on the quantum channel, i.e., $W=1$ with $V_S=10^3$. (b) Close up view of (a) where $V_0=1.5$ for heterodyne detection (solid (blue) line) crosses over with the pure state (dotted (red) line) for homodyne detection.}\label{RR_comparision}
\end{center}
\end{figure}

We can now compare homodyne detection to heterodyne detection using reverse reconciliation with, for example, an impurity of $V_0=1.5$. This is plotted in Fig.~\ref{RR_comparision}~(a). We note that after a certain value of line transmission ($\approx T>0.79$) it is better, in terms of information rates, to use heterodyne detection with a noisy input state than homodyne detection with a pure input state, c.f., Fig.~\ref{RR_comparision}~(b).

\section{Direct Reconciliation}

We now look at direct reconciliation~\cite{Grosshans2002}~($\blacktriangleright$) where Bob uses both homodyne and heterodyne detection. First though, we begin with our analysis using homodyne detection as first presented in~\cite{Weedbrook2010}.

\subsection{Homodyne Detection}\label{Sec: DR Hom}

\begin{figure}[!ht]
\begin{center}
\includegraphics[width=8cm]{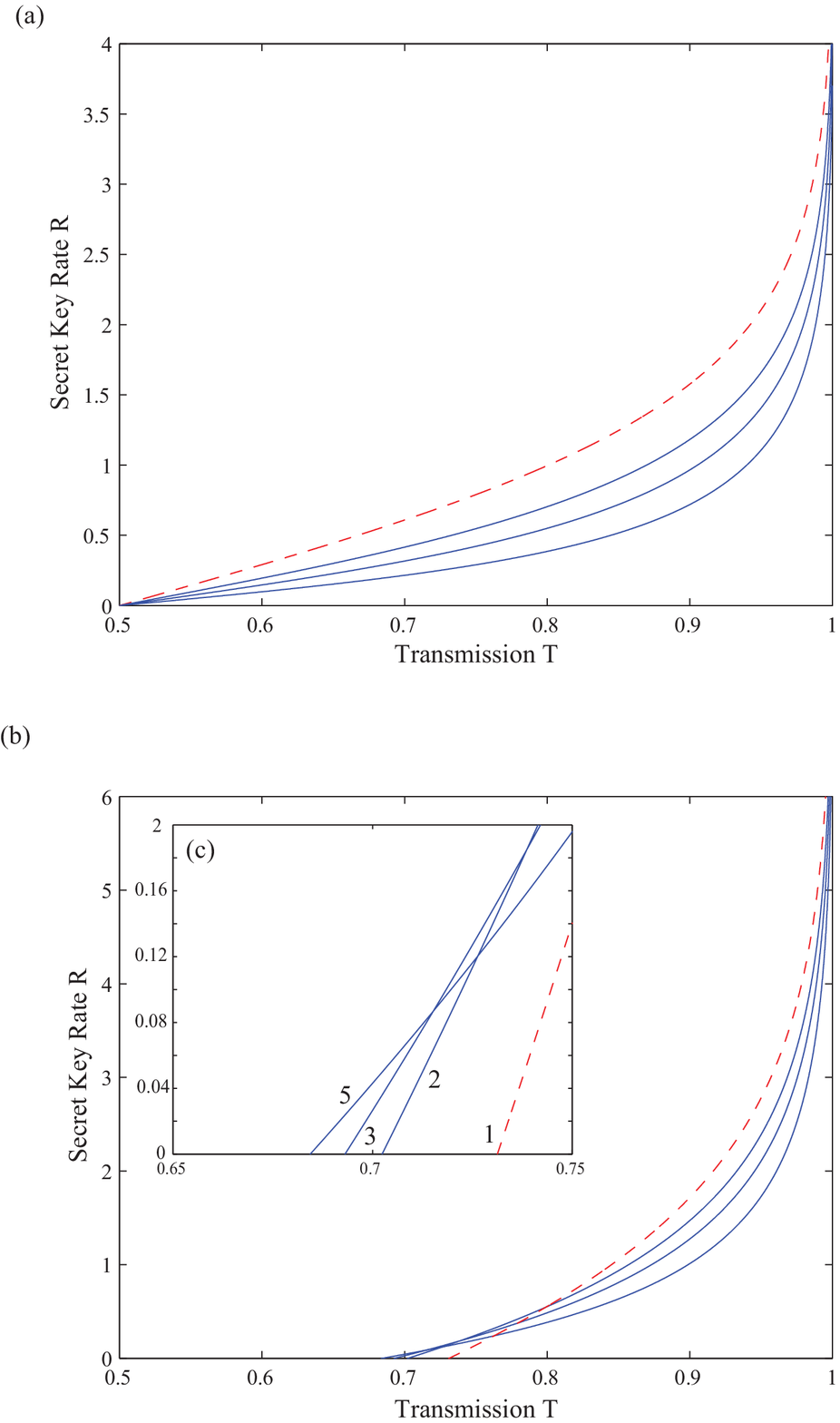}
\caption{Secret key rates for direct reconciliation using (a) homodyne detection and (b) heterodyne detection. Here the dashed (red) line is the pure mode case $V_0=1$ where the solid (blue) lines are for impurity values $V_0 =2,3,5$, from top to bottom; again using the parameters: $W=1$ and $V_S=10^3$. (c) Close up view of (b) showing that as Alice's input state becomes more and more thermal, even though the information rates are reduced, the protocol becomes more secure in terms of where the lines cross the transmission axis. The values of $V_0$ are indicated next to the respective lines.}\label{DR_Hom_all}
\end{center}
\end{figure}

The secret key rate for direct reconciliation using homodyne detection is given by
\begin{align}\label{eq: secret key rate hom DR}
R^{\blacktriangleright [Hom]} := I(X_A:X_B) - I(X_A:E),
\end{align}
where $I(X_A:X_B)$ has already been calculated in Eq.~(\ref{eq: alice and bob mutual hom RR formula}) (note the mutual information between Alice and Bob is symmetric with respect to the two reconciliation protocols). For Eve we have
\begin{align}\label{eq: alice and eve mutual info}
I(X_A:E) := S(E) - S(E|X_A),
\end{align}
where $S(E|X_A)$ is calculated from the spectrum $\boldsymbol{\nu}_{E|X_A}$ of the conditional covariance matrix $\textbf{V}_{E|X_A}$. Eve's conditional covariance matrix for homodyne detection using direct reconciliation is equal to
\begin{align}
\textbf{V}_{E|Q_A} = \textbf{V}_E (V_0,V),
\end{align}
where $\textbf{V}_E$ is defined in Eq.~(\ref{eq: eves CM}). The resulting symplectic spectrum calculated using Eq.~(\ref{eq: symplectic eigenvalues def}) is again too complicated to be written down here. However, in Fig.~\ref{DR_Hom_all}~(a) we have plotted the resulting secret key rates for various values of $V_0$. Here we see the surprising feature of direct reconciliation as first noticed in \cite{Weedbrook2010} where adding preparation noise onto the initial states does not reduce the transmission range of the protocol (despite the fact that the secret key rate is reduced).

\subsection{Heterodyne Detection}

\begin{figure}[!ht]
\begin{center}
\includegraphics[width=8cm]{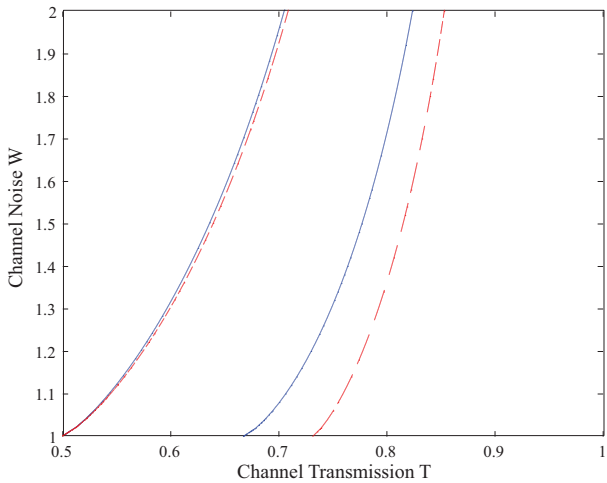}
\caption{Security threshold plots for direct reconciliation where the (red) dashed lines indicate a pure vacuum mode at Alice's preparation side and the (blue) solid lines indicate a noisy coherent state ($V_0 \gg 1$) as input. The two lines converging to $T=0.5$ are for homodyne detection while the other two lines indicate heterodyne detection. Adding preparation noise to the heterodyne detection protocol illustrates the largest improvement in security of the pair of two protocols.}\label{DR_threshold_bounds}
\end{center}
\end{figure}

In our final analysis of this section, we consider heterodyne detection using direct reconciliation. The secret key rate for direct reconciliation using homodyne detection is given by
\begin{align}
R^{\blacktriangleright [Het]} := I(X_A:X_B) - I(X_A:E),
\end{align}
where $I(X_A:X_B)$ is the same as Eq.~(\ref{eq: alice and bob mutual Het RR}). For Eve, her mutual information with Alice is defined as
\begin{align}
I(X_A:E) := S(E) - S(E|Q_A,P_A),
\end{align}
where $S(E|Q_A,P_A)$ is calculated from the spectrum $\boldsymbol{\nu}_{E|Q_A,P_A}$ of the conditional covariance matrix $\textbf{V}_{E|Q_A,P_A}$. This conditional covariance matrix is given by
\begin{align}
\textbf{V}_{E|Q_A,P_A} = \textbf{V}_E (V_0,V_0),
\end{align}
where again $\textbf{V}_E$ is defined in Eq.~(\ref{eq: eves CM}). Using Eq.~(\ref{Spectrum_simple1}) we can write the symplectic spectrum as
\begin{align}
{\bf \nu}_{E|Q_A,P_A}^{\pm} = \frac{1}{2} \Big[\sqrt{(e_1 + W)^2 - 4 T (W^2 -1)} \pm (e_1 - W)\Big].
\end{align}
The resulting secret key rates are plotted in Fig.~\ref{DR_Hom_all}~(b) for different values of initial mode impurity. As with homodyne detection, when the impurity is increased, there is no reduction in the security threshold of the protocol, only the secret key rates. However, the security threshold for heterodyne detection ($T \approx 0.73$) is higher than that of homodyne detection ($T=0.5$). Surprisingly though, by adding more and more uncertainty to the initial modes, the security threshold \textit{improves} slightly for heterodyne detection; meaning that the protocol can, at least for a small window of transmissions, tolerate slightly higher levels of loss (cf., Fig.~\ref{DR_Hom_all}~(c)). For example, for a pure vacuum as input a secure key can be generated from a transmission of $T >\approx 0.73$. However, when the initial mode is set to $V_0=5$ we have $T >\approx 0.68$. Numerically, we find that for large values of impurity $V_0 \gg 1$, the security asymptotes to $T \rightarrow 0.67$. Out of the four family of protocols studied here, this is the only protocol that exhibits such behavior. Figure~\ref{DR_threshold_bounds} contains plots of the security thresholds (where $R=0$) for both homodyne and heterodyne detection using direct reconciliation and shows the improvement in security when (unknown) preparation noise is added, with heterodyne detection offering the largest improvement.  This situation of noise improving the performance of QKD has previously been seen in the context of direct reconciliation where (pure) coherent states and homodyne detection are more robust than squeezed states and homodyne detection~\cite{Navascues2005}. Furthermore, reverse reconciliation, where Bob measures squeezed states using heterodyne detection rather than homodyne detection, also shows an enhanced robustness~\cite{Gar09,Pirandola2009}. Achieving such security robustness only works when additional noise is added to the reference point (either Alice or Bob) of the reconciliation protocol. This means Alice in direct reconciliation and Bob in reverse reconciliation.

\begin{figure}[!ht]
\begin{center}
\includegraphics[width=8cm]{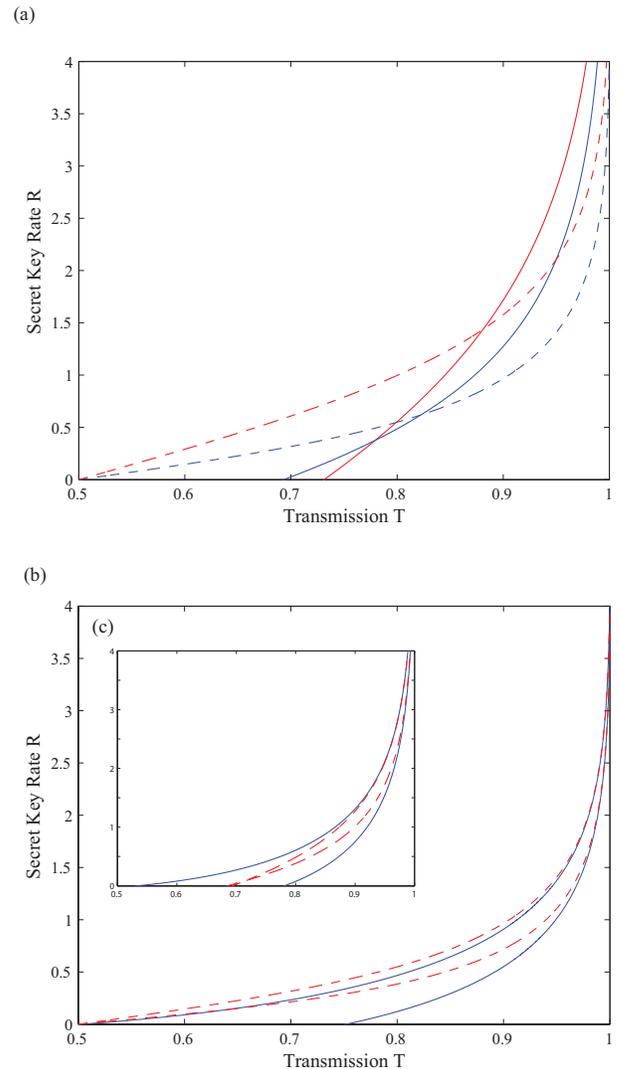}
\caption{(a) Comparison of secret key rates between direct reconciliation using homodyne detection (dashed lines) and heterodyne detection (solid lines) for $V_0 = 1$ (red lines) and $V_0 = 3$ (blue lines) and with only loss on the quantum channel, i.e., $W=1$, with $V_S=10^3$. Comparison of direct and reverse reconciliation for (b) homodyne detection and (c) heterodyne detection for a lossy channel. Here the dashed (red) lines indicate direct reconciliation whilst the solid (blue) lines indicate reverse reconciliation. In each of the cases we have plotted the impurity values of $V_0 =3$ and $5$.}\label{DR_comparision}
\end{center}
\end{figure}

As we did with the reverse reconciliation protocols, we compare homodyne detection to heterodyne detection but this time for direct reconciliation. This comparison is plotted in Fig.~\ref{DR_comparision}~(a) where we have compared the two pure vacuum modes against $V_0=3$ for both homodyne and heterodyne detection. We have also plotted a comparison between direct and reverse reconciliation for both homodyne and heterodyne detection using impurity values of $V_0 = 3$ and $5$. In the case of homodyne detection, as given in Fig.~\ref{DR_comparision}~(b), we find that direct reconciliation offers stronger security and higher information rates than reverse reconciliation for the same values of impurity. This is somewhat mirrored in the heterodyne detection scenario given in Fig.~\ref{DR_comparision}~(c), although it only becomes more apparent for values of impurity higher than $V_0=5$.

\subsection{Effect of Channel Noise}

\begin{figure}[!ht]
\begin{center}
\includegraphics[width=8cm]{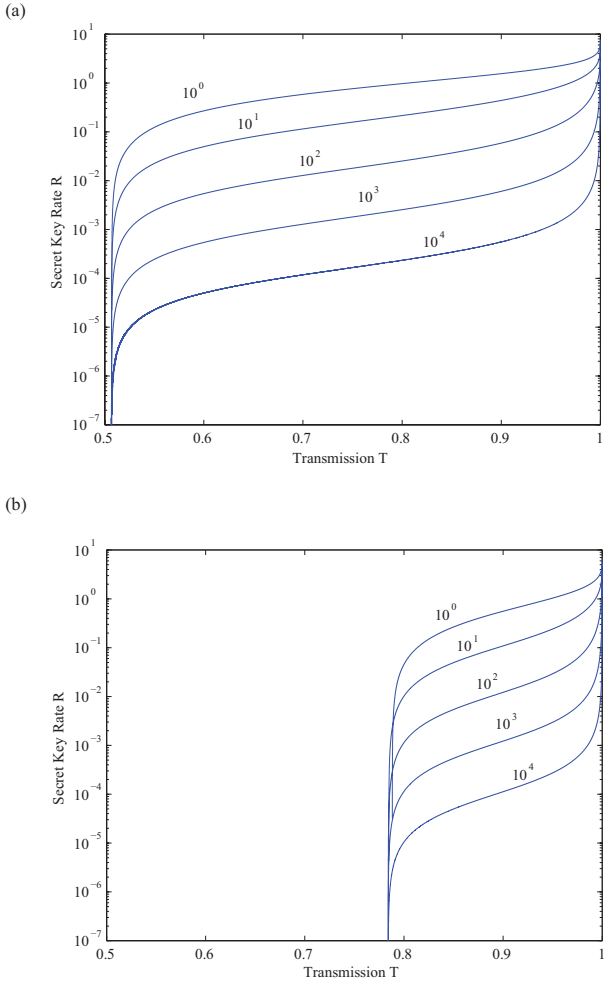}
\caption{The effect of channel noise $W$ on the secret key rates for direct reconciliation using homodyne detection for a quantum channel with noise of (a) $W=1.01$ and (b) $W=3$ for various values of $V_0$ where $V_S=10^5$. The effect of increasing the channel noise only shifts the security threshold of both plots while maintaining the same characteristic of the lossy channel~\cite{Weedbrook2010} where all values of $V_0$ converge to the same channel transmission. Such a preparation noise effect is not seen in reverse reconciliation.}\label{classical_limit_DR_Hom_noise}
\end{center}
\end{figure}

Here we consider the effect of channel noise ($W>1$) on the protocol that uses direct reconciliation with homodyne detection for larger values of preparation noise. In particular we consider preparation noises with a variance up to $10^4$. In Fig.~\ref{classical_limit_DR_Hom_noise} we plot the two cases of channel noises of $W=1.01$ and $W=3$ (Fig.~\ref{classical_limit_DR_Hom_noise}~(a) and (b), respectively). As expected both plots show a reduction in both the channel transmission and secret key rate for both channel noises. However, the characteristic where the various values of $V_0$ converge to the same channel transmission value still remains.

\section{QKD in the classical limit}

So far we have considered what happens to the four protocols (direct/reverse reconciliation using homodyne/heterodyne detection) when modest amounts (at most $V_0=5$) of preparation noise is added onto Alice's input states. In this section we consider the `classical limit', where an infinite amount of preparation noise is added, i.e., $V_0 \rightarrow \infty$, and hence, the quantum vacuum mode contribution to QKD becomes (almost) negligible. Here we consider the case of direct reconciliation (using homodyne detection) because as we have seen, reverse reconciliation (using either homodyne and heterodyne detection) does not handle preparation noise very well as the security (channel transmission) and the secret key rate deteriorate quickly for modest increases in noise. It was first shown in~\cite{Weedbrook2010} that for a pure loss channel ($W=1$), a secret key could still be established even if Alice's preparation noise was as large as $V_0=10^4$. Adding preparation noise from $V_0=1$ to $V_0=10^4$ reduced the key rate but kept the maximum transmission threshold fixed at $T=0.5$.

We now consider what happens to the secret key rate $R^{\blacktriangleright [Hom]}$ in the asymptotic limit where the preparation noise goes to infinity $V_0 \rightarrow \infty$ and the channel noise is much smaller, i.e., $W \ll V_0$. To do this we consider the 
fixed ratio
\begin{align}
\phi := \frac{V_S}{V_0} >0.
\end{align}
In our calculations we make the substitution $V_S= \phi V_0$ and take the limit $V_0 \rightarrow \infty$. We begin by first considering the mutual information between Alice and Bob as defined in Eq.~(\ref{eq: alice and bob mutual hom RR formula}). Following the recipe given above we obtain
\begin{align}
I(X_A:X_B) = \frac{1}{2} \log_2 (1 + \phi).
\end{align}
The above equation is simply Shannon's formula for the classical capacity of a single-mode communication channel with additive Gaussian noise of variance $V_0$ and input Gaussian signal $V_S$~\cite{Sha48}. We now calculate the mutual information between Eve and Alice $I(X_A:E)$ in this so-called classical limit. To do this we follow the techniques given in Sec.~IV~A. Note that Eve and Alice's mutual information is defined in Eq.~(\ref{eq: alice and eve mutual info}) and uses Eq.~(\ref{eq: von neuman entropy}) with Eq.~(\ref{eq: g of von Neumann}). However, in this asymptotic limit the symplectic eigenvalues are also very large, i.e., $\nu \gg 1$, in which case Eq.~(\ref{eq: g of von Neumann}) is simplified to
\begin{align}
g(\nu) \rightarrow g'(\nu)= \log_2 \Big(\frac{e \nu}{2}\Big) + O(\nu^{-1}).
\end{align}
Again in this asymptotic limit, the first symplectic spectrum value of Eve using Eq.~(\ref{eq: symplectic eigenvalues of Eve}) is given by
\begin{align}
\nu_{E}^+ = (1-T)V.
\end{align}
While the other symplectic eigenvalue can be calculated in the same manner to give
%
\begin{align}
\nu_{E}^- = W.
\end{align}
To calculate the conditional symplectic spectrum $\boldsymbol{\nu}_{E|X_A}$ we use Eq.~(\ref{eq: symplectic eigenvalues def}) where
\begin{align}\nonumber
\Delta = &W^2 +(V-TV+TW)(V_0 - TV_0 + TW)\\
&-2T(W^2-1),
\end{align}
and
\begin{align}\nonumber
&(\Delta^2 -  4{\rm det} {\bf V}_{E|X_A}) = (T-1)^2 [T^2 (V-W)^2 (V_0 - W)^2\\
&+ (W^2-VV_0)^2 + 2T(V-W)(W-V_0)(W^2+VV_0-2)].
\end{align}
Taking the limit as before gives, for the first symplectic eigenvalue, the following
\begin{align}
\nu_{E|X_{A}}^+ = \sqrt{1+ \phi}  (1-T) V_0   + O(V_0^{-1}).
\end{align}
Now in order to get a non-zero value for the other symplectic eigenvalue we use Eq.~(\ref{eq: variables in terms of SS}) to give
\begin{align}\nonumber
&\nu_{E|X_{A}}^- =\\
&(1-T)^{-1} \sqrt{\frac{(T +VW-TVW)(T+V_0W-TV_0W)}{V V_0}}.
\end{align}
Using the above asymptotic formulas with Eq.~(\ref{eq: secret key rate hom DR}), we find that a positive secret key rate exists only when
\begin{align}
R^{\blacktriangleright [Hom]} = \log_2 \Big[\frac{(\sqrt{1+\phi}) \nu_{E|X_{A}}^+\nu_{E|X_{A}}^-}{\nu_{E}^+\nu_{E}^-} \Big] >0.
\end{align}
The above expression can be simplified to
%
\begin{align}\label{eq: simplified expression of limit condition}
\log_2 \Big[\frac{[T+V W(1-T)][T+V_0 W(1-T)]}{V V_0 W^2 (1-T)^2}\Big] >0.
\end{align}
For a finite information rate we therefore require the following inequality to be true
\begin{align}
\frac{[T+V W(1-T)][T+V_0 W(1-T)]}{V V_0 W^2 (1-T)^2} > 1.
\end{align}
Algebraically, we find that the above inequality is always satisfied for our required conditions of $1/2<T<1$, $V_0>1$, $V_S>1$ and $W>1$. Therefore we have shown that in the asymptotic limit where $V_0 \rightarrow \infty$, $V_S = \phi V_0$, and $W \ll V_0$, any value of preparation noise can be added onto the initial quantum states used by Alice and a secret key can still be achieved, albeit with a very small, but still finite, key rate. This happens as long as the transmission of the channel is greater than a half.

\section{Quantum cryptography at various electromagnetic wavelengths}

\begin{figure}[!ht]
\begin{center}
\includegraphics[width=8cm]{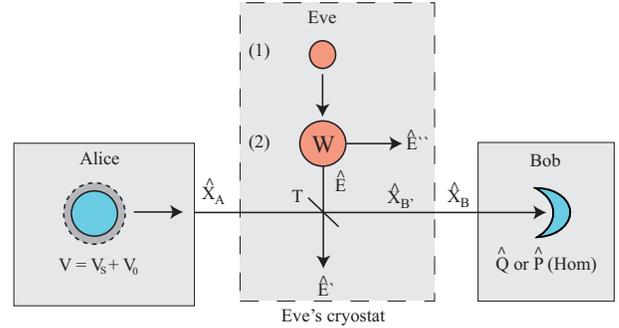}
\caption{Schematic of a continuous-variable QKD protocol performed at different wavelengths of the electromagnetic spectrum. Alice sends modes at a particular fixed wavelength to Bob who measures the incoming modes using homodyne (Hom) detection. Eve's attack consists of using a cryostat which is used to cool down her thermal modes to produce pure vacuum modes (1). The second step (2) involves implementing the entangling cloner attack. Finally, Eve adds known noise onto the modes $\hat{X}_{B'}$ she sends to Bob. This is to create a thermal state in order to match the level of the variance of the radiation of the environment, effectively covering her tracks.}\label{fig_1_way_2_microwave}
\end{center}
\end{figure}

It is interesting to consider that one possible application of the results from the previous two sections is continuous-variable QKD over different wavelengths of the electromagnetic spectrum. The reason why the previous analyzes would be useful for such an application is that the average photon number is dependent on the wavelength of the signals sent. Typically, QKD experiments~\cite{Sca09} are performed at telecom wavelengths of $1550$~nm where the average photon number at room temperature is very low~($\bar{n} \sim 10^{-14}$). However, when one moves away from this wavelength and down into the infrared, the modes become more thermal. In Sec.~IV we determined that direct reconciliation is significantly more robust against preparation noise than reverse reconciliation and is therefore better suited to our analysis of QKD at various wavelengths. We consider a simply model where Alice sends Bob thermal states at a particular wavelength and Bob uses a (perfect efficiency) homodyne detector that is unaffected by the thermal radiation. Note that if Bob employed heterodyne detection the additional unit of shot noise vacuum from the heterodyne detector would also be thermal and need to be taken into account. For Eve's attack, as with the previous sections we
assume she performs a collective Gaussian attack, but this time with a difference. Eve's ancilla modes, which she interacts with Alice's incoming modes (where the interaction is typically modeled using a Gaussian beam splitter), are also thermal (for the same reason Alice's are). In order to combat this Eve performs her entire attack inside a cryostat, see Fig.~\ref{fig_1_way_2_microwave}. In preparation for her attack Eve's first step (1) is to cool down her thermal ancilla modes so as to approximate pure vacuum modes. In the second step (2) she performs a collective Gaussian attack via the entangling cloner attack. Then, before sending the mode $\hat{X}_{B'}$ onto Bob, she randomly modulates $\hat{X}_{B'}$ to create (from Alice and Bob's perspective) a thermal state. The variance of this thermal state is chosen to be equal to the variance of the environmental noise, so that Eve covers her tracks. The key point here is that by modulating her modes Eve has effectively added \textit{known} noise to her ancilla modes.

\begin{figure}[!ht]
\begin{center}
\includegraphics[width=8cm]{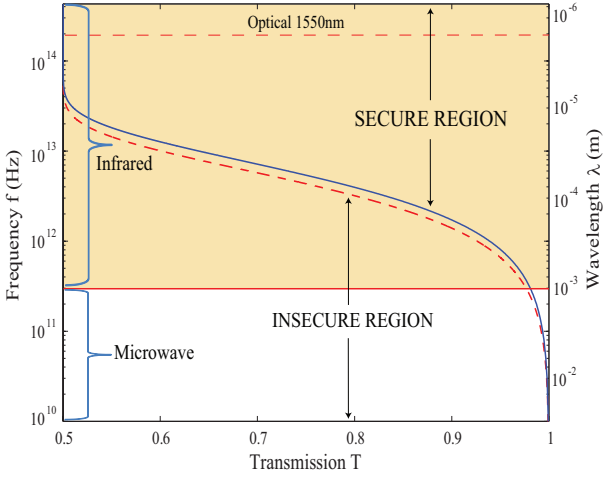}
\caption{Security of continuous-variable QKD as a function of channel transmission $T$ at various wavelengths of the electromagnetic spectrum at room temperature $\tau = 300$~K. Beginning at the infrared spectrum ($430$~THz) and down into the microwave spectrum (starting from $300$~GHz) where $V_S=10^8$. The solid (blue) line is the secure region derived against a collective Gaussian attack. The dotted (red) line corresponds to an entanglement-breaking channel where Eve performs an intercept-resend attack. In such a situation no secure key can be synthesized.}\label{spectrum3}
\end{center}
\end{figure}

To begin the analysis we need to calculate the variance of a mode at a specific wavelength. To do this we note that we can write the average photon number $\bar{n}$ in terms of the quadrature variance $V$ as
\begin{align}\label{eq: average photon number}
\bar{n} = \langle \hat{a}^{\dagger} \hat{a} \rangle = \frac{1}{2} (V-1) \Longrightarrow V = 2 \bar{n}+1,
\end{align}
where $\hat{a} = (\hat{Q} + i \hat{P})/2$ and $\hat{a}^{\dagger}$ are the annihilation and creations operators, respectively, and we have also symmetrized both quadratures, i.e., $V := V(\hat{Q}) = V(\hat{P})$. Now the average photon number for a single mode is equal to~\cite{Ger05}
\begin{align}\label{eq: average photon number2}
\bar{n} = \frac{1}{\exp(h f / k_B \tau) - 1},
\end{align}
where $\tau$ is the temperature, $f$ is the frequency of the mode, $h$ is Planck's constant and $k_B$ is Boltzmann's constant. Using the techniques from Sec.~\ref{Sec: DR Hom} we can calculate the regions where continuous-variable QKD is secure as a function of the frequency~(wavelength) and channel transmission. This is plotted in Fig.~\ref{spectrum3} where areas of security correspond to $R >0$ where again $R$ is the secret key rate. We see that regions of security exist over various wavelength values from optical ($1550$~nm) into the infrared and down into the microwave region. We note that in the original paper~\cite{Weedbrook2010}, where continuous-variable QKD at various frequencies was first investigated, a bound was derived that underestimated the security threshold. The new tigher bound given in Fig.~\ref{spectrum3} improves upon the previous bound by having higher levels of security.

It is instructive to consider a loss limit (or transmission threshold) for QKD at various wavelengths. It is known~\cite{Namiki2004} that a loss limit exists when considering channel noise for continuous-variable QKD. This bound corresponds to Eve performing an intercept-resend attack which destroys any quantum correlations between Alice and Bob and thus the possibility of generating a secure key~\cite{Curty2004}. In order to avoid an entanglement-breaking channel we demand that the equivalent noise of the quantum channel $\chi$ cannot exceed one unit of shot-noise, i.e., $\chi<1$~\cite{Grosshans2002}. Since $\chi= W(1-T)/T$, the security condition becomes
\begin{align}\label{eq: EPR breaking}
W < \frac{T}{1-T}.
\end{align}
We can rewrite the above equation in terms of a secure bound on the required frequency as a function of channel transmission. Using the fact that $W = 2 \bar{n} +1$ with Eqs.~(\ref{eq: average photon number2}) and (\ref{eq: EPR breaking}) we can show that we require
\begin{align}
f > -\alpha \ln (2T-1),
\end{align}
where $\alpha = k_B \tau/h$. This curve is plotted as the dotted (red) line in Fig.~\ref{spectrum3} and gives a lower bound in security.

\begin{figure}[!ht]
\begin{center}
\includegraphics[width=8cm]{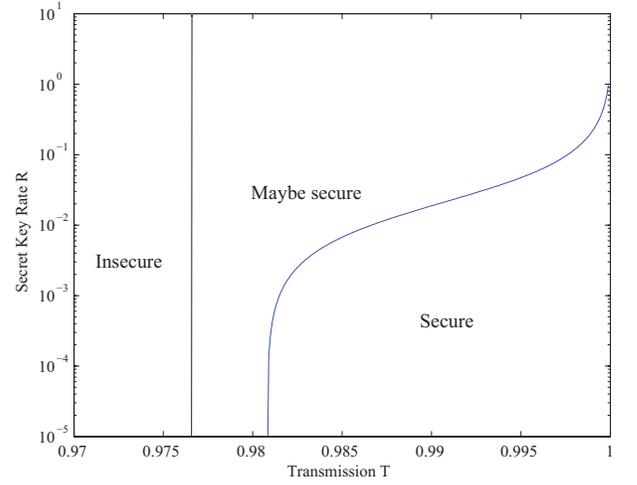}
\caption{Security of microwave quantum cryptography. Here we consider the upper end of the microwave spectrum, i.e., $V_0=41.66$ (300~GHz) using the direct reconciliation protocol and homodyne detection, where $W=41.66$ and $V_S=10^8$. The insecure region corresponds to the entanglement-breaking channel where no secure key can be created; whilst a region in between the secure and insecure region exists where it might be secure but as yet no known protocol exists. For example, secure protocols could be developed which are based on more
complex strategies in terms of classical
communications and post-processing.}\label{DR_microwave3}
\end{center}
\end{figure}
\begin{figure}[!ht]
\begin{center}
\includegraphics[width=8cm]{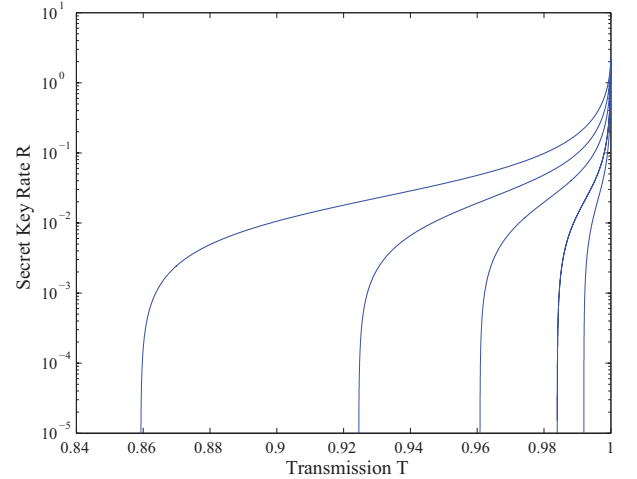}
\caption{How channel noise $W$ affects the security of thermal state QKD. Here we have $V_0=41.66$ using direct reconciliation and homodyne detection with $V_S=10^3$. From left to right: $W=5,10,20,50,100$. After only modest increases in channel noise the security of the protocol reduces rapidly.}\label{Microwaves_excess_channel_noise}
\end{center}
\end{figure}

\subsection{Discussion: Feasibility of Microwave QKD}

Here we consider the possibility of using QKD at the microwave frequency. The microwave frequency is ubiquitous as a communication wavelength in today's technologies, ranging from cell phones to short-range devices such as Wi-Fi and Bluetooth. The fact that small regions of security exist in the microwave regime is initially quite surprising due to the presence of large amounts of background noise. We consider the microwave frequency from $300$~GHz~($1$~mm) to $1$~GHz~($30$~cm). Using Eq.~(\ref{eq: average photon number2}) for Alice's initial modes we find that this corresponds to a range of variances from $V_0=41.66$ to $V_0=1.25 \times 10^4$, respectively. In Fig.~\ref{DR_microwave3} we plot the case where $V_0 = 41.66$ (i.e., $300$~GHz) and where the noise on Eve's mode is also $W = 41.66$. We see that a secure key can only be generated when the transmission is higher than $T \approx 0.981$. Here the straight vertical line distinguishing the insecure and secure regions is the entanglement-breaking region as given by Eq.~(\ref{eq: EPR breaking}), i.e., $T>W/(1+W)=0.9766$. For the 1~GHz frequency, numerically we only start getting positive key rates when the channel reflection (i.e., loss) is on the order of $1-T \approx 10^{-5}$ giving a key rate on the order of $R \approx 10^{-6}$. Although the secure region is very small, the practical required distances are also very small. Such a short-range QKD scheme, unlike the typical long-range QKD protocols, could potentially be ideal for such devices as Bluetooth (maximum distance of $\sim 10$~m) and Wi-Fi ($\sim 75$~m). Also a secure quantum version of Near Field Communication~(NFC)~\cite{NFC} would be an ideal application as the range with which these microwave devices operate over is $\sim 10$~cm. However, in such a situation what actually constitutes Alice's and Bob's stations becomes blurred. We point out that the dominate factor in terms of the limited range in security, is the channel noise $W$ and, as we have seen from the results of the previous section, not the preparation noise. The effect of channel noise on the security of thermal state QKD is plotted in Fig.~\ref{Microwaves_excess_channel_noise}. Here we assume $V_0=41.66$ and see that after only a small increase in channel noise (i.e., $W=5$) one can only generate a secure key after $T \approx 0.86$. Therefore a continuous-variable QKD protocol that is able to tolerate large amounts of preparation \textit{and} channel noise is required, in order to make microwave QKD feasible.

Another possible platform for microwave QKD is using discrete variables, e.g., the BB84 protocol~\cite{BB84}. The preparation and detection of photons at the microwave frequency is an active field of experimental research in cavity quantum electrodynamics~\cite{Varcoe2000,Raimond2001,Schuster2007,Houck2007}. However, such experiments do not involve the propagation of microwave photons over free space. Although, even if the technology allowed the efficient generation and detection of single microwave photons over free space, the fundamental problem exists where Bob would not be able to distinguish the photons that originated from Alice to those which came from the surrounding environment - both are indistinguishable.

\section{Conclusion}

In conclusion, we have considered continuous-variable quantum key distribution from the perspective of Alice using thermal Gaussian states as her initial cryptographic resource, instead of the usual pure Gaussian states. The case of direct reconciliation and homodyne detection was first analyzed in~\cite{Weedbrook2010} and we have extended these results here to include both direct and reverse reconciliation for the case of heterodyne detection. We showed that an improved robustness to channel noise can be achieved when preparation noise is added in the case of direct reconciliation using heterodyne detection. In~\cite{Weedbrook2010} it was shown that direct reconciliation does not suffer any loss in security when preparation noise is added (although the secret key rate does decrease as a function of preparation noise), even when the variance of the initial thermal states was as large as $10^4$ times that of the pure vacuum. We significantly improved upon this result by showing that direct reconciliation can tolerate any amount of preparation noise, provided the channel noise is much less than the preparation noise. Finally, we derived an upper bound related to an entanglement-breaking eavesdropping attack for quantum key distribution at various electromagnetic wavelengths and ended with a discussion on the feasibility of microwave quantum key distribution.

\acknowledgments

C.~W. would like to thank Bing Qi, Hoi-Kwong Lo, Li Qian, Travis Humble and Nathan Walk for discussions. We would also like to thank Nathan Walk, Carlo Ottaviani and Xiang-Chun Ma for pointing out a mistake in a previous version. C.~W. acknowledges support from the Ontario postdoctoral fellowship program,
CQIQC postdoctoral fellowship program, CIFAR, Canada Research Chair program,
NSERC, and QuantumWorks. S. P. acknowledges support from EPSRC under grant
No. EP/J00796X/1 (HIPERCOM) and the European Union
under grant agreement No. MOIF-CT-2006-039703. T.~C.~R acknowledges support from the Australian Research Council.

\end{document}